# Graphene-based thermal repeater

Ming-Jian He[1,2], Hong Qi[1,2,*], Ya-Tao Ren[1,2], Yi-Jun Zhao[1], and Mauro Antezza[3,4,**]

1 School of Energy Science and Engineering, Harbin Institute of Technology, Harbin 150001, P. R. China

2 Key Laboratory of Aerospace Thermophysics, Ministry of Industry and Information Technology, Harbin 150001, P. R. China

3 Laboratoire Charles Coulomb (L2C), UMR 5221 CNRS-Université de Montpellier, F-34095 Montpellier, France

4 Institut Universitaire de France, 1 rue Descartes, F-75231 Paris, France

*Corresponding authors: Email: qihong@hit.edu.cn (H. Qi), mauro.antezza@umontpellier.fr (M. Antezza)

**Abstract:** In this letter, we have demonstrated the possibility to efficiently relay the radiative heat flux between two nanoparticles by opening a smooth channel for heat transfer. By coating the nanoparticles with a silica shell and modifying the substrate with multilayered graphene sheets respectively, the localized phonon polaritons excited near the nanoparticles can couple with the multiple surface plasmon polaritons near the substrate to realize the heat relay in the long distance. The heat transfer can be enhanced by more than six orders of magnitude and the relay distance can be as high as 35 times in the far-field regime. The work may provide a way to realize the energy modulation or thermal communications especially in long distance.

Since the prediction by Polder and Van Hove [1] that the heat exchange between two objects can be much higher than the blackbody limit at nanoscale separations, numerous nanoscale thermal devices [2-8] and unique phenomena [9, 10] are proposed. Thermal repeater is a device that can help the heat propagate further, acting as a relay. In other words, the key function of it is amplifying the radiative heat transfer (RHT) in long distance. When the two objects are separated at long distance, the RHT decreases dramatically as a result of the deterioration of the evanescent wave channel. Based on that, Dong et al. [11] proposed a method to delay the deterioration between two nanoparticles by placing a substrate near the two nanoparticles to form a channel for propagating surface waves. However, its performance dramatically degrades for metal nanoparticles, due to lack of coupled modes between the nanoparticles and the dielectric substrate. By coating monolayer graphene on the substrate, the performance of the long-distance energy-exchange has been improved [12]. However, the amplification is still limited for the reason that the graphene sheet can only couple with the substrate to form surface modes, but cannot greatly tailor the interplay between the nanoparticles and substrate. Moreover, the amplification is sorely confined near the interface of the substrate, due to the surface characteristics of graphene surface plasmon polaritons (SPPs). The lack of the connecting between the particles and the substrate, and the nature of SPPs both restrict the relay effect.

Graphene exhibits intriguing electronic properties including the presence of strongly confined SPPs which can be excited to transport energy in photonic channels [13, 14]. This has been exploited for a more efficient RHT between hybrid periodic structures [15] and between nanoparticles [16]. Recently, by using multilayer structures, the RHT between NPs has been shown to be further increased thanks to a resonant coupling between the substrate surface modes and the NPs resonances [17]. In addition, some unique phenomena are observed in multilayered graphene system [15, 18, 19], and one of the most important features is the multiple SPPs excited by the multilayered graphene. The effect provides the possibility to strengthen the SPPs near the interface of the substrate, in other words, as surface modes, the SPPs can spread further from the interface. Considering the lack of the channel between metal particles and the substrate, the core/shell configuration of nanoparticles can be utilized to realize the modifying of the particles [20, 21]. Inspired by the above, we introduce a concept of a thermal repeater by modifying the particles with shell and embedding the substrate with multilayered graphene sheets, to form strongly coupled modes. It should be mentioned that the shell together with the graphene sheets are the key components of the repeater, which can introduce the strong interactions between nanoparticles and substrate, and support much stronger surface modes than monolayer graphene.

To start, let us consider two silica-encapsulated gold nanoparticles located near a silica substrate with a distance $h$ from the interface, and the separation distance between the nanoparticles is denoted as $d$ in FIG. 1. For the encapsulated nanoparticles, the radius of the core is $R_{in}$ and the outside radius of the particles is $R_{in}+t$. Based on the framework of Mie-Lorenz theory, the electric and magnetic polarizabilities of the encapsulated nanoparticles can be modified and given as [22]

$$\alpha_E^{(0)}(\omega) = 4\pi R^3 \frac{(\varepsilon_s - \varepsilon_h)(\varepsilon_c + 2\varepsilon_s) + f(\varepsilon_c - \varepsilon_s)(\varepsilon_h + 2\varepsilon_s)}{(\varepsilon_s + 2\varepsilon_h)(\varepsilon_c + 2\varepsilon_s) + f(\varepsilon_c - \varepsilon_s)(2\varepsilon_s - 2\varepsilon_h)} \tag{1}$$

$$\alpha_M^{(0)}(\omega) = \frac{2\pi}{15} R^3 \left(\frac{\omega R}{c}\right)^2 \left[\frac{\varepsilon_s - \varepsilon_h + f^{5/3}(\varepsilon_c - \varepsilon_s)}{\varepsilon_h}\right] \tag{2}$$

where $R$, $\varepsilon_c$ and $\varepsilon_s$ are the radius of the nanoparticles, the dielectric function of the core and shell respectively. $\varepsilon_h$ is taken as 1 when the particles are in vacuum. $f = (R_{in}/R_{out})^3$ is the core volume fraction of the encapsulated nanoparticles. It is easy to show that, for $f = 1$ or $\varepsilon_c = \varepsilon_s = \varepsilon_{Au}$, the polarizabilities reduce to those of the bare Au nanoparticles. In this letter, to avoid the effect of total volume of the particles on RHT, we compare the results of bare Au nanoparticles with radius 10 nm to silica-encapsulated gold nanoparticles with $R_{in} = 7$ nm and $t = 3$ nm. To guarantee the validity of the dipolar approximation, we will limit our calculations to the particle-surface distance $h$ and particle-particle distance $d$ both larger than 100 nm [12, 17]. It is worth mentioning that in experiments, controlled growth of silica shells from about 1.7 to 4.3 nm is realized with gold nanoparticles which have diameters smaller than 20 nm [23]. In order to make the study more meaningful for experiments and realistic devices, the Drude critical point model for the dielectric function of gold is used [24]. Moreover, considering the finite size effects in small core radius and small shell thickness, the dielectric is modified to a radius-dependent one and the detailed formulas can be found in Ref. [25]. The dielectric properties of silica is based on Ref. [26]. In the fluctuation-dissipation theorem describing dipole fluctuations, the polarizability needs to be modified by [27]

$$\chi(\omega) = \text{Im}\left[\alpha(\omega)\right] - \frac{k_0^3}{6\pi}|\alpha(\omega)|^2 \tag{3}$$

where $\alpha(\omega) = \alpha^{(0)}(\omega) / \left[1 - i\omega^3 \alpha^{(0)}(\omega) / (6\pi c^3)\right]$ is the dressed polarizability with the radiation correction and $k_0 = \omega/c$.

Now we consider the RHT in the system. The whole system, including the nanoparticles and the substrate are assumed to be thermalized at $T$=300 K, then one of the nanoparticles is heated up to $T+\Delta T$. This leads to a radiative heat flux $\varphi$ between two nanoparticles and then the radiative heat transfer conductance (RHTC) is defined to quantitatively evaluate the RHT between the two nanoparticles

$$C = \lim_{\Delta T \to 0} \frac{\varphi}{\Delta T} = 4 \int_0^{+\infty} \frac{\hbar \omega^5}{2\pi c^4} \cdot \frac{\partial n(\omega, T)}{\partial T} \chi^2 \mathrm{Tr}\left(\mathbf{G} \mathbf{G}^*\right) d\omega \tag{4}$$

where $n(\omega, T) = \left[\exp(\frac{\hbar \omega}{k_{\mathrm{B}} T}) - 1\right]^{-1}$ is the Bose-Einstein distribution and * denotes conjugate transpose. $\mathbf{G}$ is the dyadic Green tensor which is composed of two parts, a vacuum contribution $\mathbf{G}^{(0)}$ and a scattering contribution $\mathbf{G}^{(\mathrm{sc})}$, accounting for the direct particle-particle exchange and the particle-interface interplay. The two parts read respectively as

$$\mathbf{G}_{\mathrm{E}}^{(0)} = \mathbf{G}_{\mathrm{M}}^{(0)} = \frac{e^{ik_0 d}}{4\pi k_0^2 d^3} \begin{pmatrix} a & 0 & 0 \\ 0 & b & 0 \\ 0 & 0 & b \end{pmatrix} \tag{5-a}$$

$$\mathbf{G}_{\mathrm{E}}^{(\mathrm{sc})} = \int_0^{+\infty} \frac{ik e^{2ik_z z}}{4\pi k_0^2 k_z} \left(R_s \mathbf{S} + R_p \mathbf{P}\right) dk \tag{5-b}$$

where $a = 2 - 2ik_0 d$, $b = k_0^2 d^2 + ik_0 d - 1$. $k$ and $k_z = \sqrt{k_0^2 - k^2}$ are the parallel and perpendicular wave-vectors. The scattering contribution $\mathbf{G}_{\mathrm{E}}^{(\mathrm{sc})}$ in Eq. (5-b) is the component caused by electric dipoles and magnetic part $\mathbf{G}_{\mathrm{M}}^{(\mathrm{sc})}$ caused by magnetic dipoles can be obtained by exchanging the Fresnel coefficients $R_s$ and $R_p$. More details including the matrixes $\mathbf{S}$ and $\mathbf{P}$ can be found in Refs. [11, 12]. As for the conductivity of graphene, it can be written as a sum of an intraband and an interband contribution, respectively, given by [28]

$$\sigma_{\mathrm{intra}}(\omega) = \frac{2ie^2 k_{\mathrm{B}} T}{(\omega + i\tau^{-1})\pi \hbar^2} \ln\left[2\cosh\left(\frac{\mu}{2k_{\mathrm{B}} T}\right)\right] \tag{6-b}$$

$$\sigma_{\mathrm{inter}}(\omega) = \frac{e^2}{4\hbar} \left[G(\frac{\hbar \omega}{2}) + \frac{4i\hbar \omega}{\pi} \int_0^{+\infty} \frac{G(\eta) - G(\frac{\hbar \omega}{2})}{(\hbar \omega)^2 - 4\eta^2} d\eta\right] \tag{6-b}$$

where $G(x)$=sinh($x/k_{\mathrm{B}}T$)/[cosh($\mu/k_{\mathrm{B}}T$)+cosh($x/k_{\mathrm{B}}T$)], $e$ and $k_{\mathrm{B}}$ are elementary charge and Boltzmann constant, respectively. As can be seen from Eq. (5-b), the graphene sheet on the substrate, comes into play by modifying the surface characteristics [12]. The silica substrate provides surface phonon polaritons which can couple with the graphene SPPs [8]. For the silica shell on the nanoparticles, the effect of it cannot be found in the formulas of the Green tensor, while the polarizabilities are modified because of it. We note that in Eq. (4), the Green tensor and the polarizabilities, which reflect the characteristics of the interface and particles respectively, jointly contribute to the RHTC.

The main point of the letter is to show how the simultaneous presence of embedded-multilayered graphene and silica shell, and in particular of the coupled multiple SPPs and localized phonon polaritons, is able to act as a thermal repeater in long-distance energy-exchange. To this aim, in FIGs. 1(b) and 1(c), the amplification factor $\eta_1 = C_j/C_1$ and $\eta_3 = C_j/C_3$, viewed as the performance of the whole repeater and the pure effect of graphene, is given as a function of $d$ when $h$=100 nm, where $j$ denotes the index for the five cases: (i) $j$=1-bare substrate and nanoparticles, (ii) $j$=2-graphene-coated substrate and bare particles, (iii) $j$=3-bare substrate and encapsulated nanoparticles, (iv) $j$=4-graphene-coated substrate and encapsulated nanoparticles and (v) $j$=5-substrate embedded with $N$-layer graphene sheets and encapsulated nanoparticles. For cases (i)-(iv), the substrates are semi-infinite, while for case (v), the thickness of the substrate is $H$=100 nm and the multilayered graphene are evenly embedded. The chemical potential of graphene is set as 0.5 eV throughout the letter. The inset in FIG. 1(b), $C_2/C_1$, together with $C_4/C_3$ in FIG. 1(c), reveal that only by covering substrate with monolayer graphene, the RHT cannot be enhanced greatly. However, according to $C_3/C_1$ and $C_4/C_1$, with the maximum $2.50 \times 10^4$ and $6.83 \times 10^4$, we can show that the RHTC can be enhanced by more than four orders of magnitude when the nanoparticles are encapsulated. In addition, when the shell and multilayered graphene work together, the $C_5/C_1$ and $C_5/C_3$ lines reveal that the heat transfer can be enhanced by another two orders of magnitude. The phenomenon indicates that by modifying the substrate with multilayered graphene sheets, the effect of graphene can come into play. The more layers of embedded graphene sheets also endow the RHT with larger separation distance to be enhanced. It should be pointed out that, the RHTC is the same for case (i) at $d$=1.44 $\mu$m and case (iv) $N$=10 at $d$=50 $\mu$m, which means the thermal repeater can realize about 35 times relay distance.

In Eq. (4), the RHTC can be written in two kinds of forms, $C=C_E+C_M$ and $C=C^{(0,0)}+C^{(0,sc)}+C^{(sc,sc)}$, where $C_E$, $C_M$, $C^{(0,0)}$ , and $C^{(0,sc)}+C^{(sc,sc)}$ account for the contribution of electric dipoles, magnetic dipoles, the direct particle-particle flux and the scattering flux based on the substrate [12]. To get insight into the physical mechanism responsible for the heat relay, we present the electric RHTC ratio $\eta_E = C_E/C$ and scattering RHTC ratio $\eta_S = C/C^{(0,0)}$ in FIGs. 2(a) and 2(b), respectively. As is known, for silica/graphene and gold, the heat transports mainly in electric and magnetic channels respectively [29]. In combination with the physical mechanism of the particle-interface interplay defined by $\mathbf{G}^{(sc)}$, the effect of the substrate can be displayed by $\eta_S$. Then the important role that graphene sheet and silica shell play can be clearly evaluated and distinguished by analyzing the $\eta_E$ and $\eta_S$ simultaneously. The near-zero $\eta_{E,1}$ and low $\eta_{S,1}$ indicate that, for case (i) without silica shell or graphene, the RHT is dominated by magnetic dipoles of gold, and transports mainly in the direct particle-particle channel, without benefit of the substrate. Compared with case (i), when the substrate is coated with monolayer graphene, the increased $\eta_{E,2}$ and $\eta_{S,2}$ near 2 $\mu$m imply that graphene sheet starts to work and assists to exchange energy. Nevertheless, the limited $\eta_{S,2}$ means that the effect of graphene is still tiny, due to weakness of graphene SPPs at the large particle-interface distance $d=100$ nm and lacking channel between metal particles and dielectric substrate. When the gold nanoparticles are encapsulated with silica shell, the characteristics of the particles are completely modified, and the electric dipoles replace the magnetic dipoles to dominate the RHT, which can be seen from the near-unity $\eta_E$ for case (iii), (iv) and (v). The opening of the electric channel for particles also induces another effect, that is, improving the possibility of exchanging energy by interacting with the graphene. The increasing trend from $\eta_{S,1}$ to $\eta_{S,3}$, and from $\eta_{S,2}$ to $\eta_{S,4}$ confirms the effect. While, the scattering ratio $\eta_S$ for cases (iv) is still limited and the maximum is approximately ten, which means that the effect of the graphene is not fully exploited. The low $C_4/C_3$ in FIG. 1(c) can be well explained by the fact that the monolayer graphene on the substrate loses efficacy at the large particle-interface distance, even though the electric channel has been fully exploited. When the substrate is embedded with multilayered graphene sheets, i.e., case (v), the scattering ratio increases dramatically, and the maximum for $N=5$ reaches 229 at $d=2.31$ $\mu$m, which is approximately 20 times that of case (iv). Considering the results in FIGs. 1(b) and 1(c) for case (v), we can confirm that the embedded-multilayered graphene sheets are the key component of the system to realize long-distance energy-exchange, by providing a smooth channel for particles to transport heat in scattering manner.

To get an insight into the near-unity electric ratio $\eta_E$ that causes approximately four orders of magnitudes of enhancement for RHT, we plot in FIG. (3a) the direct particle-particle RHTC ratios between encapsulated and bare Au nanoparticles. According to the mechanisms of the system governed by Eqs. (3)-(5), only the characteristics of the particles, i.e., the electric and magnetic polarizabilities are responsible for the direct particle-particle RHT. Therefore, the ratios are uniform between the cases (i)/(ii) and cases (iii)-(v). An interesting phenomenon has been observed that the ratio at $d$=1 $\mu$m initiate to increase and then approximately doubles beyond $d$=6 $\mu$m. The increasing trend of the particle-particle RHTC ratios also provides more possibility for particles to exchange heat in electric channel at longer separation distance. Therefore, the graphene sheets can benefit from the effect and then the enhancement caused by graphene is able to work at longer distance. The interplay between the particles and the plasmonic system is governed by the local density of electromagnetic states (LDOS). The LDOS enhancement, i.e., the LDOS near the nanoparticles, $\rho^{\mathrm{E}}$, relative to the LDOS in free space, $\rho_0^{\mathrm{E}}$, is given with an integral over frequency in FIG. (3b). The detailed formulas for calculating the LDOS for isolated nanoparticles can be found in Ref. [30]. For bare Au nanoparticle, the LDOS enhancement is tiny and confined near the particles. While, for encapsulated nanoparticles, the LDOS enhancement can be enhanced by several orders of magnitude near the particles and spreads further. Moreover, the enhancement is strong even for very high filling factor of Au $f$=0.8, which means only 0.7 nm of shell, let alone the considered size in the letter, $R_{\mathrm{in}}$=7 nm and $t$=3 nm with $f$=0.343.

To understand the large scattering ratio $\eta_{\mathrm{S}}$ at $d$=4.33 $\mu$m caused by multilayered graphene, which is the key performance of graphene to enhance the RHT, we next focus on the spectral RHTC for $N$=1, 5 and 10 in FIG. 4(a). The RHTC without graphene is also given in the figure to make out the pure effect of graphene. The total RHTC and scattering ratio $\eta_{\mathrm{S}}$ without graphene are $9.35\times10^{-25}$ JK$^{-1}$rad$^{-1}$ and 0.95 at the specific distance, while the RHTC and $\eta_{\mathrm{S}}$ for $N$=1, 5 and 10 multilayered graphene system are $1.00\times10^{-23}$ JK$^{-1}$rad$^{-1}$/10, $7.81\times10^{-23}$ JK$^{-1}$rad$^{-1}$/79, and $1.47\times10^{-22}$ JK$^{-1}$rad$^{-1}$/153 respectively. The adding of graphene induces the dramatic increase in scattering ratios and hence the total RHTC, especially for multilayered graphene sheets. Moreover, to demonstrate the effect of graphene clearly, we show in FIG. 4(b) the Fresnel coefficients $R_p$. According to the dielectric function of silica, it has two frequency bands for phonon polaritons, and they are labeled in the figure. We can show that as the number of graphene sheets increases, there are more bands of graphene SPPs modes and then form multiple SPPs, which can greatly couple with silica phonon polaritons to enhance the RHT. To characterize the combined effect of the embedded-multilayered graphene sheet and the encapsulated particles, the electric field energy density distribution $u_{\mathrm{e}}$ [11] are calculated at $\omega$=2.2$\times10^{14}$ rad/s and displayed in FIG. 4(c). The $x$ axis means the interface of the substrate and the two red points at $x/d=\pm0.5$ are the positions of the nanoparticles. The electric field energy density distribution for bare Au nanoparticles without graphene, encapsulated particles without graphene, $N$=1 and $N$=5 multilayered graphene system are given from the top to the bottom. Similar to the results of LDOS enhancement in FIG. 3(b), the electric field energy density for bare Au particles are also confined near the particles and can hardly interact with the substrate. Interestingly, after the nanoparticles are encapsulated, the electric field energy density greatly strengthens, which is mainly attributed to the localized phonon polaritons exited near the particles. When the graphene is added in the system, graphene SPPs are excited near the interface and take part in exchanging energy by surface modes. For $N$=5, multiple SPPs are dramatically excited and the electric field energy density fills up the region near the interface of the substrate between the two particles. The results demonstrate that the localized phonon polaritons of the encapsulated particles can interact with the multiple SPPs of the graphene sheets to form strongly coupled modes, opening a smooth channel in the long-distance energy-exchange. Up to now, the underlying mechanism of the heat relay and amplification is explored and demonstrated clearly.

In conclusion, we have theoretically proposed a thermal repeater by coupling modes of graphene, and it allows us to increase the RHT as high as six orders of magnitude even in the long distance without any additional source of energy. By distinguishing the contributions between nanoparticles and graphene, electric and magnetic dipoles, we have found that the amplification and relay are realized by the strong coupling between localized phonon polaritons excited near the encapsulated nanoparticles and multiple SPPs modes of graphene. In this letter, we have limited to the simplest shape of the nanoparticles, however, the mechanism and concept proposed here could be easily applied to much more complex configurations. The letter also provides the possibility to realize the thermal communications for transmitting information by photons of thermal objects rather than electrons of currents.

The supports of this work by the National Natural Science Foundation of China (No. 51976044, 51806047) are gratefully acknowledged. The authors thank J. Dong for fruitful discussions. M. A. acknowledges support from the Institute Universitaire de France, Paris, France (UE).

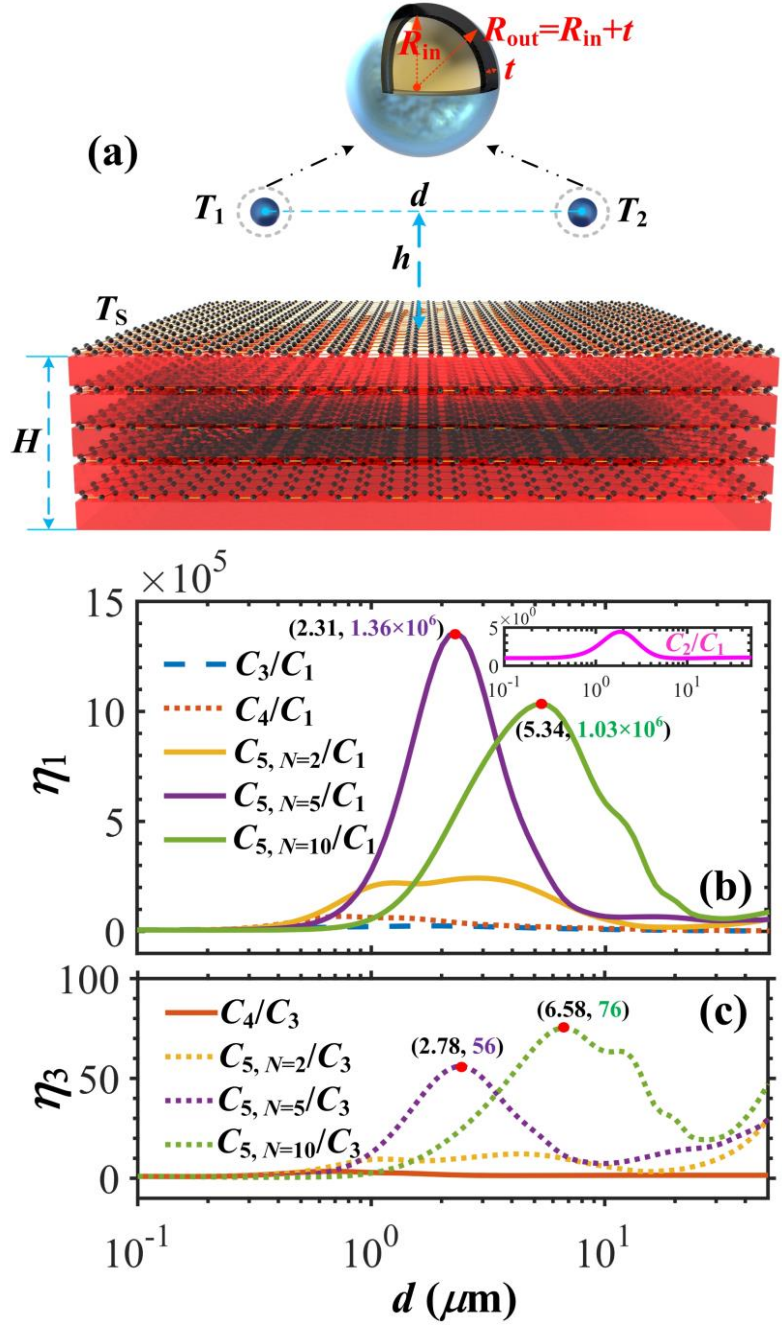

FIG. 1 (a) Schematic of RHT between two encapsulated Au nanoparticles located near a substrate embedded with multilayered graphene sheets. The RHTC enhancement ratios compared to (b) case (i) and (c) case (iii).

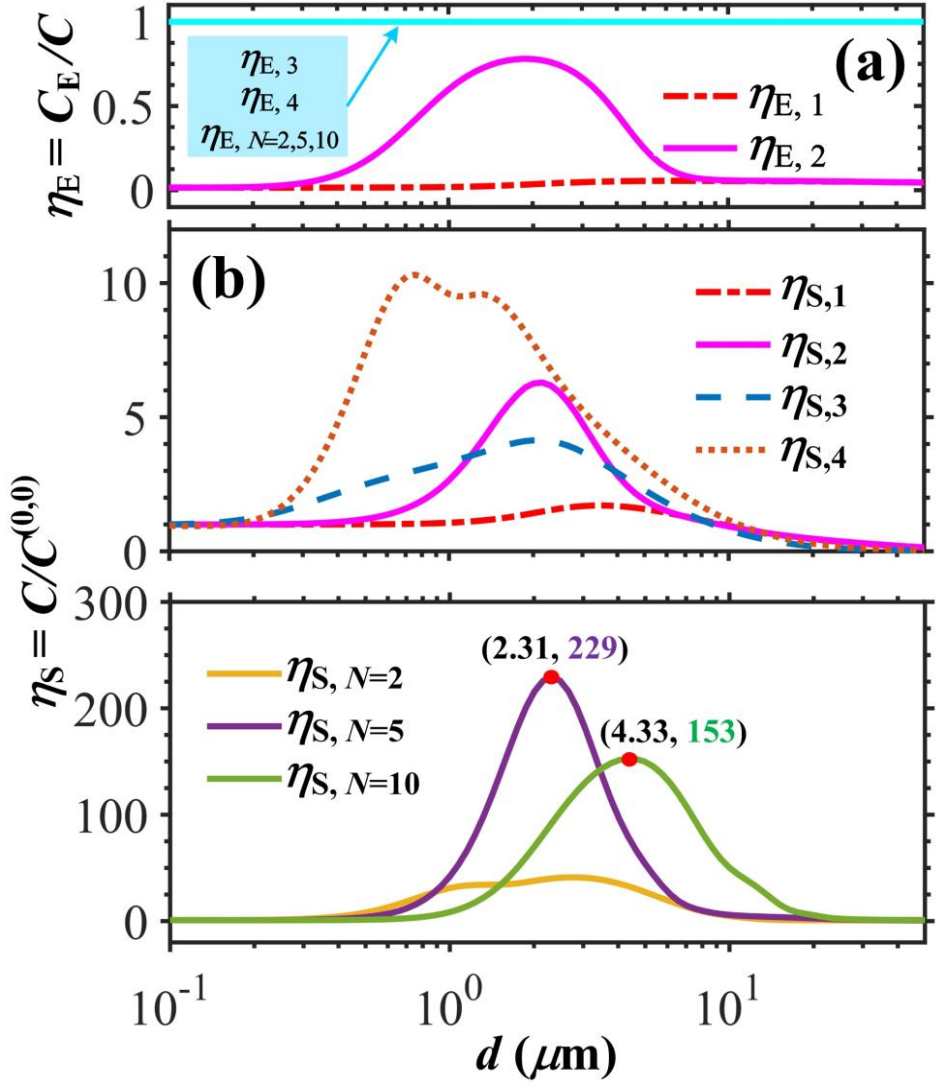

FIG. 2 For different $d$, (a) the electric RHTC ratio $\eta_E = C_E/C$ and (b) scattering RHTC ratio $\eta_S = C/C^{(0,0)}$ for different cases.

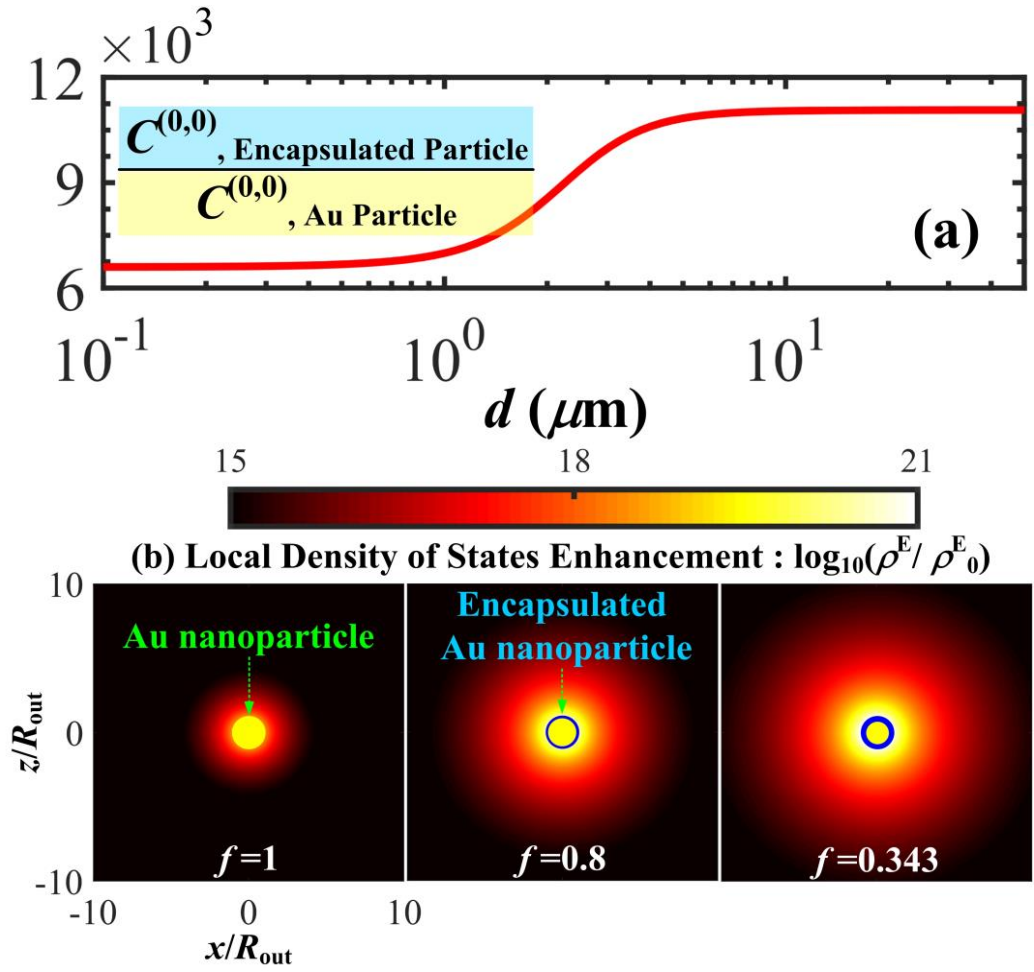

FIG. 3 (a) The direct particle-particle RHTC ratios between encapsulated and bare Au nanoparticles. (b) The local density of electromagnetic states enhancement near the isolated nanoparticles.

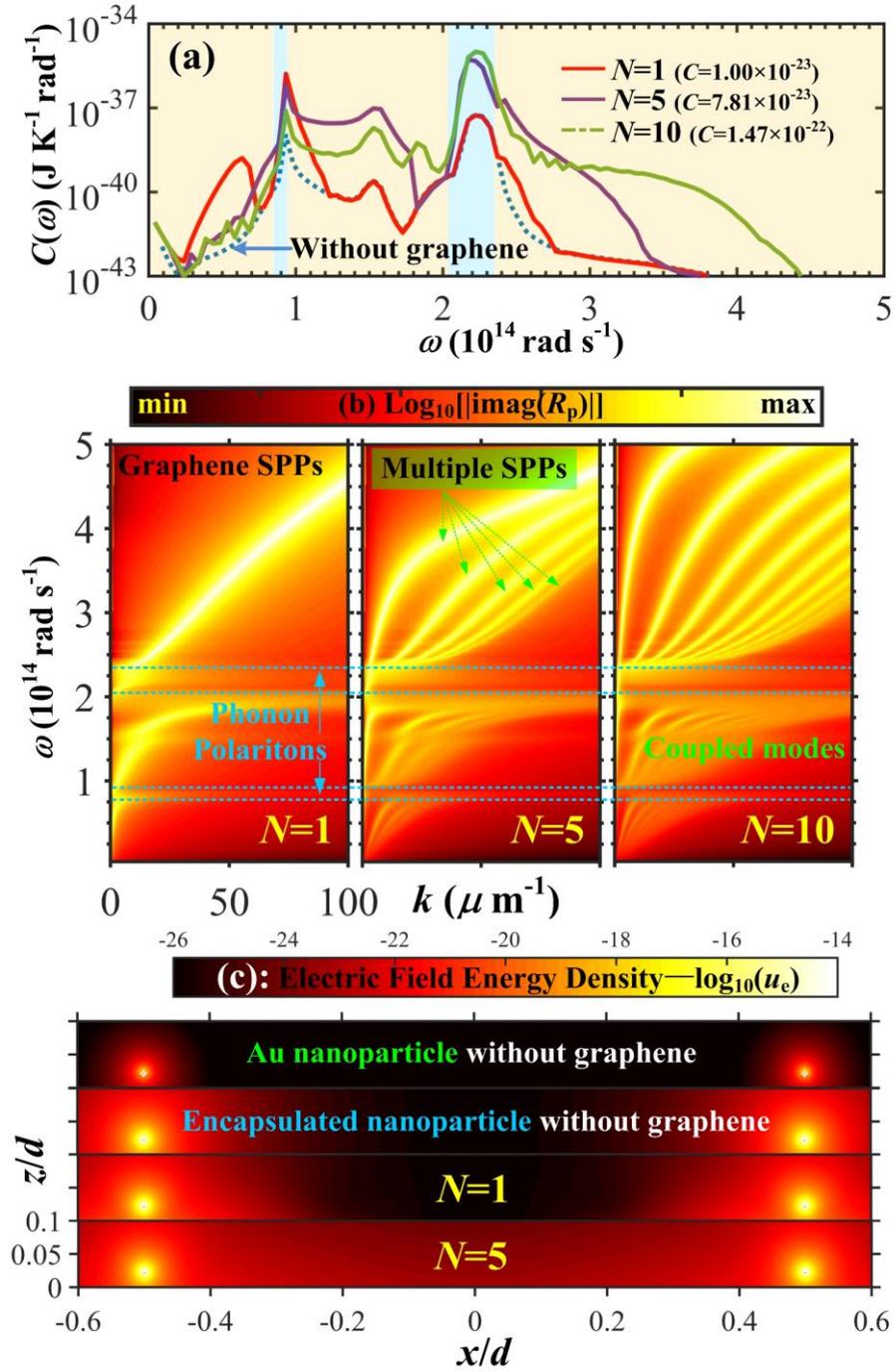

FIG. 4 (a) The spectral RHTC at $d$=4.33 $\mu$m. (b) The Fresnel coefficients $R_p$ for the substrate embedded with multilayered graphene sheets. (c) The electric field energy density distribution for different cases near the interface.